\documentclass[10pt]{article} %

\usepackage{amssymb}            %
\usepackage{mathtools}          %
\usepackage{mathrsfs}           %
\usepackage{graphicx}           %
\usepackage{subcaption}         %
\usepackage[space]{grffile}     %
\usepackage{url}                %
\usepackage{lipsum}             %
\usepackage{multirow}
\usepackage{makecell}
\usepackage{array}
\usepackage{wrapfig}
\usepackage{blindtext}
\usepackage{booktabs} 

\usepackage{hyperref}
\usepackage{natbib}
\usepackage{authblk}

\usepackage{cleveref} %

\crefname{figure}{Figure}{Figures}
\Crefname{figure}{Figure}{Figures}

\title{Quantitative Resilience Modeling for Autonomous Cyber Defense}

\author{Xavier Cadet\textsuperscript{1,$\dagger$}, Simona Boboila\textsuperscript{2,$\dagger$}, Edward Koh\textsuperscript{1}, Peter Chin\textsuperscript{1}, Alina Oprea\textsuperscript{2}}

\affil{
$^{1}$Dartmouth College\\
$^{2}$Northeastern University%

\par
$^\dagger$ Equal contribution
}
\date{}

\begin{document}

\maketitle

\begin{abstract}
Cyber resilience is the ability of a system to recover from an attack with minimal impact on system operations.
However, characterizing a network's resilience under a cyber attack is challenging, as there are no formal definitions of resilience applicable to diverse network topologies and attack patterns.
In this work, we propose a quantifiable formulation of resilience that considers multiple defender operational goals, the criticality of various network resources for daily operations, and provides interpretability to security operators about their system's resilience under attack. We evaluate our approach within the CybORG environment, a reinforcement learning (RL) framework for autonomous cyber defense, analyzing trade-offs between resilience, costs, and prioritization of operational goals.
Furthermore, we introduce methods to aggregate resilience metrics across time-variable attack patterns and multiple network topologies, comprehensively characterizing system resilience.
Using insights gained from our resilience metrics, we design RL autonomous defensive agents and compare them against several heuristic baselines, showing that proactive network hardening techniques and prompt recovery of compromised machines are critical for effective cyber defenses.
\end{abstract}

\section{Introduction}
Cyber attacks can cause massive economic damage to an organization, lead to loss of information and privacy, and adversely affect all aspects of our society. 
Although techniques for defending cyber networks against attacks have been studied for a long time, rigorous methods to evaluate the impact of attacks on a system and its operations are generally lacking~\citep{Fleming2021}.
Cyber resilience, the ability of a system to resist and recover from a compromise, has been gaining attention as a key property of systems in cyber defense~\citep{Kott2021, Linkov2023, Weisman2025}. 
However, quantifying cyber resilience is challenging, as it involves trade-offs between different security and operational objectives and their associated costs. 

A resilient system must be able to absorb and mitigate the effect of an attack and adapt quickly to new threats. With recent developments in autonomous cyber operations, reinforcement learning (RL) provides the appropriate framework to design adaptive and optimal defense strategies.  
Autonomous solutions have the potential to reduce the burden on security operators when dealing with large search spaces over computer network features that contain vulnerabilities and entry points of attacks. Typically, RL-based autonomous defenses are evaluated by their cumulative returns~\citep{vyas2023automated, Mcdonald2024, hammar2024optimal}, but their impact on resilience in cyber networks has not been studied. 
 
In this paper, we define and evaluate new resilience metrics for cyber networks that generalize to multiple network topologies and attack patterns,  provide interpretability to security operators, and support multiple resilience objectives as prioritized by defenders. We use the insights provided by resilience metrics to develop new RL-based defensive agents that incorporate both proactive actions and prompt recovery of detected threats. 
In more detail, our main contributions are as follows:
\begin{itemize}
\item \textbf{Quantifying resilience:} We provide a quantifiable formulation of resilience that takes into account the operational goals of the defender (such as confidentiality, integrity, and availability) and the criticality of various network resources. 
We evaluated our metric using an operational workflow simulated in CybORG~\citep{cyborg_acd_2021}, a state-of-the-art cybersecurity RL environment. Our code is publicly available~\footnote{ \url{https://github.com/xcadet/CyberResilience}}.

\item \textbf{Attacks evolving over time:} We show how to evaluate resilience over time to gain insights about evolving attack patterns and system defenses, such as: Did the attack ever ramp up or was the defense able to absorb the compromise? How long did the system take to recover? Which defenses provide better resilience and faster response? 

\item \textbf{Balancing operational goals and costs:} We show empirically how security operators can assess and balance the resilience of their network based on operational priorities and costs.  We measure resilience in various situations of interest, such as when the availability of resources is prioritized over other objectives to provide uninterrupted service.

\item \textbf{Aggregation across attack patterns and topologies:} We show how our resilience metric can be aggregated over multiple attack patterns and multiple network topologies to provide a comprehensive evaluation of the resilience of the system in various settings.

\item \textbf{Resilient RL defense strategies:} We develop new PPO-based blue agents with resilience in mind. Our agents learn proactive network hardening strategies (such as deploying decoys on hosts to fend attackers) and reactive strategies (such as promptly restoring compromised machines to limit the attacker's movement through the network). We show that our RL agents are significantly more resilient than other heuristic agents across a wide range of attacks and network topologies.

\end{itemize}

\section{Prior Work}
\label{sec:prior}
Before taking a closer look at related research, it is worth noting that resilience has been extensively studied in various disciplines, including engineering, biology, and economics. \citet{HOSSEINI201647} undertake a review of almost 150 research articles on quantifying resilience in several fields.
In Table~\ref{tab:relwork}, we present the most relevant papers on cyber resilience.
During the last decade, several studies have looked at resilience assurances for critical infrastructure, such as electrical power plants~\citep{Francis2014AMA}, chemical plants~\citep{Rieger2014}, or isothermal reactors~\citep{Segovia2020}.  
These systems are usually modeled mathematically using linear equations based on the stability evolution of the specific physical process, a formulation that is orthogonal to our study. 
 
\citet{Fleming2021} recognized the importance of a systematic and rigorous method to manage the complexity of resilience
and developed the mission-aware cybersecurity framework. Similarly, \citet{Beling2021} proposed the Framework for Operational Resilience in Engineering and System Test (FOREST), a methodology to assess how well the resilience solution discovers and responds to attacks. These frameworks offer valuable guidelines, but without concrete mathematical formulation or quantitative tests of system resilience.
The basis for the assessment of cyber resilience in the literature is a time-dependent system performance function, $F(t)$, represented as a transition curve of system performance~\citep{Fang2016, Kott2018, Linkov2023}. 
In this representation, a more resilient system would exhibit a greater area under the curve (AUC). 
Resilience is therefore defined as the functionality averaged over the time of the mission.
\citet{Kott2021} point out that such a generic definition of resilience is insufficient. 
In order to provide a viable response consisting of identifying the threat, containing it, and recovering from the disruption, it is necessary to define and quantify functionality with respect to operational goals. 

The use of RL as a feedback mechanism for designing resilient systems has seen a surge in interest in recent years \citep{HUANG2022, Ligo2021, zhao2025}.  RL policies learn to choose the actions that optimally improve their expected return, but defining and measuring the resilience of the system remains a challenge. 
The work of \citet{Weisman2025} is one of the very few experimental studies that uses a simulated testbed to collect measurements of resilience-relevant metrics, namely the fuel efficiency and speed of a truck under attack. 
Closer to our setting of interest, cyber networks, \citet{wiebe2023} use the CybORG simulation framework to evaluate the amount of compromise in a network under attack. In this scenario, the attacker's goal is to restrict the availability of services and affect the confidentiality and integrity of data. However, the authors do not study the connection between these metrics and network resilience.

In this paper, we provide a formal definition of resilience for cyber networks under attack, which prioritizes the defender objectives and captures the attack time evolution.
To the best of our knowledge, we are the first to propose a quantifiable resilience metric in the cyber domain and use this metric to perform an in-depth comparative analysis of various defenses to achieve system resilience.

\begin{table}[t!]
   \caption{Related work on cyber resilience. }
    \begin{center}
        \begin{tabular}{>{\raggedright}p{2.5cm}llll>{\raggedright\arraybackslash}p{3cm}}\hline
            \textbf{Paper} &\textbf{\thead{Qualitative\\discussion}}
            &\textbf{\thead{Mathematical\\formulation}} &\textbf{\thead{Quantitative\\evaluation}}  &\textbf{RL} &\textbf{\thead{Objectives\\of interest\\for resilience}}
            \\ \hline 
            &&&&&\\
            Our work &\checkmark &\checkmark &\checkmark &\checkmark &confidentiality, availability, integrity \\\hline
            \citet{Weisman2025} &\checkmark &\checkmark &\checkmark &\checkmark &fuel efficiency of trucks\\\hline
            \citet{wiebe2023} & & &\checkmark &\checkmark &confidentiality, availability, integrity \\ \hline
            \citet{HUANG2022, zhao2025, Ligo2021} &\checkmark & & &\checkmark &functionality\\\hline
            \citet{ Linkov2023, Kott2018} &\checkmark &\checkmark & & &functionality\\ \hline
            \citet{Fleming2021} &\checkmark & & & &mission goals\\ \hline
            \citet{Fang2016} &\checkmark &\checkmark &\checkmark & &network link repair time\\\hline
            \citet{Segovia2020} &\checkmark &\checkmark &\checkmark & &operating pressure (isothermal reactor)\\\hline
            \citet{Francis2014AMA} &\checkmark &\checkmark &\checkmark & &number of customers receiving electric power\\\hline
            \citet{Rieger2014} &\checkmark &\checkmark &\checkmark & &product quality and waste (chemical plant)\\ \hline
        \end{tabular}
    \end{center}
    \label{tab:relwork}
\end{table}

\section{Problem Statement}
\label{sec:problem}

\begin{figure}[th]
    \centering
\includegraphics[width=0.85\linewidth]{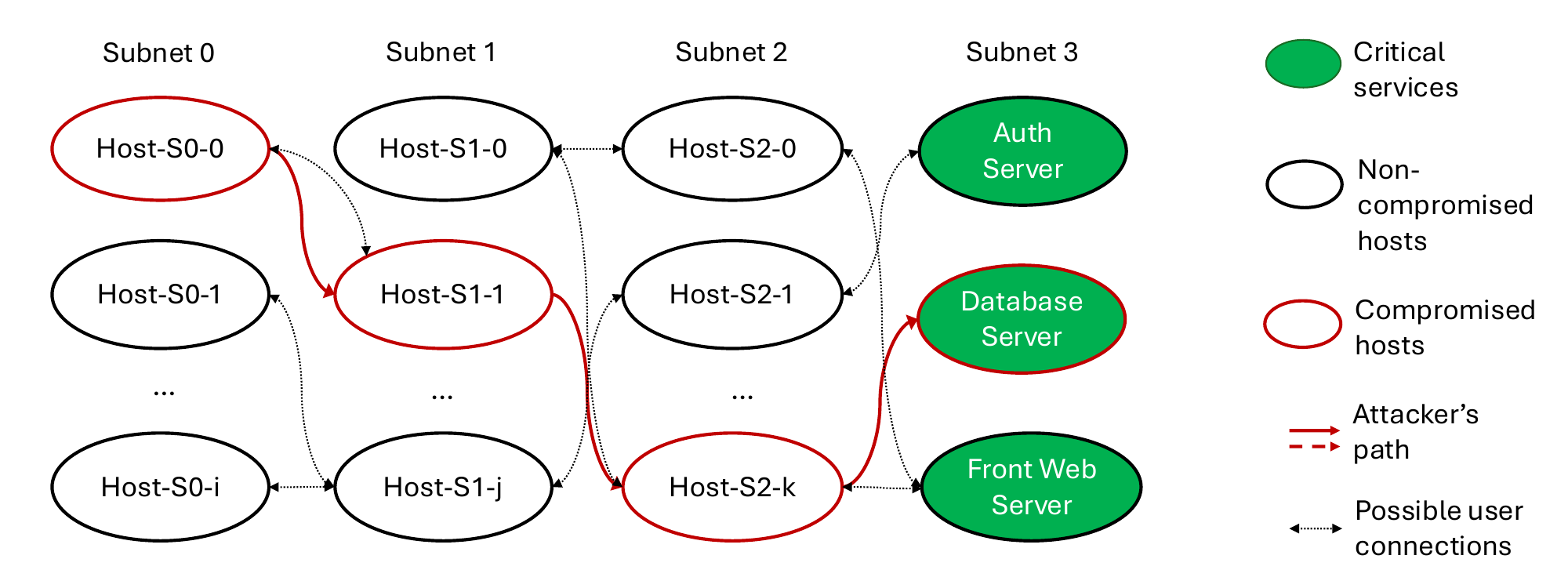}
    \caption{Topology of a cyber network, consisting of four subnets with a variable number of user machines and three critical servers for authentication, database and front web interface. The attacker's goal is to gain access to sensitive information (Confidentiality objective). The attacker establishes foothold in the network by compromising Host-S0-0 in Subnet 0, then moves laterally by compromising Host-S1-1 and Host-S2-k in Subnets 1 and 2, and finally compromises  the Database server.}
    \label{fig:network}
\end{figure}

Cyber networks are private network infrastructures of an organization designed to connect and manage devices, servers and applications. Cyber networks consist of multiple sub-networks (or subnets) to optimize performance, security, and management of resources. Examples of subnets are: client subnets including host devices such as desktops and laptops, and server subnets dedicated for critical enterprise servers, such as authentication, application, and database servers. An example of a cyber network topology is given in Figure~\ref{fig:network}, which includes three client subnets (Subnets 0, 1, and 2), and one server subnet (Subnet 3). In cyber defense, the defender's goal is to maintain the network operations, even when faced with unforeseen attacks. In particular, user and application workflows must remain operational and ensure that network resources, applications, and users interact efficiently and securely to complete their regular tasks.  We consider a case study workflow of an employee payroll system, in which employees connect to the web front end, log in using authentication credentials to submit their working hours, and retrieve data from the database server (e.g., payslips). 

\paragraph{\textbf{Adversarial objectives.}}
A cyber attack attempts to exploit network vulnerabilities and compromise host or server machines on the network to achieve specific adversarial objectives, such as:

\begin{enumerate}
    \item \textbf{Confidentiality}: The attacker obtains access to sensitive data, such as employee records that include private personal information or confidential financial documents.  
    \item \textbf{Availability}:  The attacker prevents users from achieving their operational goals by stopping an important service or overloading critical paths in the network. For instance, employees might not be able to submit their time sheets if the Database server is offline, or are logged out from important organization services if the Authentication server is not responsive. 
    \item \textbf{Integrity}: The attacker is interested in modifying data stored on a host or server, such as the company's financial records. 
\end{enumerate}

Cyber attacks consist of multiple stages over time, with an example shown in Figure~\ref{fig:network}. Typically an attack starts with establishing foothold in the network by compromising a particular host, and then propagates through the network to get to the target server. The figure shows a red path in the network from the initial compromised host to the Database server, for an adversary interested in exfiltrating employee records from the Database server (Confidentiality objective). 

\paragraph{\textbf{RL-based defenses.}} 
Recently, there has been an increasing interest in automating cyber defense strategies using RL-based agents~\citep{wiebe2023,hammar2024optimal}. To model the interaction between attackers and defenders, we use a state-of-the-art RL cybersecurity environment, 
CybORG~\citep{cyborg_acd_2021, kiely2023autonomous, cage_challenge_2_announcement}. The RL game is modeled as a partially observable Markov decision process (POMDP), a special class of MDP where the agent cannot directly observe the underlying state~\citep{Oliehoek2016ACI}.  
The attacker and defender take actions at each time step to advance the attack or implement a defensive measure. Both agents are randomized and use probabilistic policies. 

The \textit{red agent} (attacker) scans the network looking for vulnerable hosts or servers to exploit. Once it is able to create a user session on a vulnerable machine, the red agent attempts to gain root access and disrupt normal operations by performing an Impact action that targets and compromises critical services. Red agents obtain a reward if they successfully impact a host or server, and the reward value depends on the compromised machine's criticality.  

The \textit{blue agent} (defender) monitors and protects the network through a series of actions such as: analyze a host looking for malware files; start a decoy service on a host to monitor adversarial activity; remove suspicious processes from a host; restore the host to an earlier clean state. The observation space of the blue agents contains information about each host in the network, including the presence of incoming and outgoing scanning activity, and whether red sessions have been detected on a host. Blue agents obtain negative rewards if the adversary impacts a host or server, or if they perform an expensive host restore operation. Blue agents could be heuristic-based or trained with RL methods to maximize their cumulative return over episodes. 

\paragraph{\textbf{Problem definition and goals.}}
In this work, we seek to quantify the extent to which different defenses provide resilience to a cyber network during emerging attacks. Our main goal is to formally define and evaluate network resilience metrics for a quantitative assessment of system operations across time-evolving attacks and various network topologies. Resilience metrics should have the following properties: 

\begin{enumerate}

\item[P1]
\textbf{Aggregation across settings:}  
Offer a quantifiable mathematical formulation that enables the measurement of resilience at different levels of aggregation, over multiple attacks and network topologies.

\item[P2] 
\textbf{Temporal evolution:}  
Capture the temporal evolution of resilience as the attack progresses, providing interpretable insights to security operators about the resilience of a system during a cyber attack. 

\item[P3] 
\textbf{Prioritization of objectives:}
Allow defenders to prioritize multiple objectives such as confidentiality, availability and integrity, and certain services, according to their operational goals. 

\item[P4]  
\textbf{Comparison of defenses:}
Enable comparison of autonomous defenses in terms of their resilience in a repeatable and verifiable way.  
\end{enumerate}

As discussed in Table~\ref{tab:relwork}, none of the existing papers introduces resilience metrics that satisfy all these properties. While the work of \citet{wiebe2023} is the only one considering the operational goals of confidentiality, integrity and availability in cyber environments, they do not have a mathematical formulation of network resilience.  Our work aims to fill this gap in the literature. 

\section{Methodology}

We provide a quantitative formulation of resilience for a fixed attack and network topology in \Cref{sec:quant_res}, after which we discuss temporal considerations in \Cref{sec:temporal_considerations} and introduce a case study in \Cref{sec:case_study}. Finally, we discuss several RL agents for cyber defense in \Cref{sec:rl_defenses} motivated by resilience insights. 

\subsection{Quantitative Formulation of  Resilience}
\label{sec:quant_res}

Network resilience is the ability to recover from an attack with minimal impact on user and application workflows. Normal operations are dependent on critical servers that provide essential services, and these servers are usually the target of adversaries. In security scenarios, attackers are performing actions that impact or compromise critical servers over time, while defenders aim to restore these compromised services. The resilience metric aims to measure how much the impact operations on each critical asset affect the overall network resilience, according to the defender's operational goals. 

In this section, we define the resilience metric in the context of a given network topology and red agent $R$. 
By fixing the topology and the attacker's strategy, we can isolate and examine the properties of the resilience metric for different blue agents, under the same system configuration and specific threat.
Later in Section~\ref{sec:summarization} we show how the metric can be extended across topologies and attack patterns. We define the resilience metric over a time interval $\Delta t$, motivated by property P2 (ability to capture evolution patterns).
We denote by $N_j(t)$  the indicator variable of a successful adversarial impact on the critical service $j$ at time $t$, 
such that $N_j(t) = 1$ if the attacker's action disrupted the service, and $N_j(t) = 0$ if not. 
Furthermore, $cost(i,j)$ is the cost of disruption that affects the operational goal $i$ due to impact on service $j$. 

We propose a definition for resilience drop  within time interval $\Delta t$ as a weighted score that balances the defender's operational goals:
\begin{align}
\Delta R (B, R, \Delta t) = \sum_{i \in Op\_Goals} w_i \sum_{j \in Assets} \left(\sum_{t \in \Delta t} N_j(t) \right) \times  cost(i, j),
\label{eq:res}
\end{align}
\noindent where $\sum_i w_i=1$, $R$ if the
red agent's attack strategy and $B$ is the blue agent's defense strategy.

To account for property P3 in our problem definition (Section~\ref{sec:problem}), defenders can prioritize operational goals (e.g., confidentiality, availability, integrity) by selecting weights $w_i$ and can also assign different weights to critical assets $j$ for operational goals $i$  by varying $cost(i,j)$. Thus, our definition provides flexibility and can be tailored to the defender's operational goals. 

\subsection{Temporal considerations} 
\label{sec:temporal_considerations}

An important consideration for the resilience metric in  Equation~\ref{eq:res}  is the granularity of the time interval over which we measure the resilience drop. A $\Delta t$ equal to the duration of the entire game is too coarse, since quantifying $\Delta R$ on the total number of impacts loses information about the recovery process. In contrast, a single step $\Delta t$ is too fine-grained, as only one impact operation can happen during each time step.

\begin{figure}
    \centering
        \includegraphics[width=0.45\textwidth]{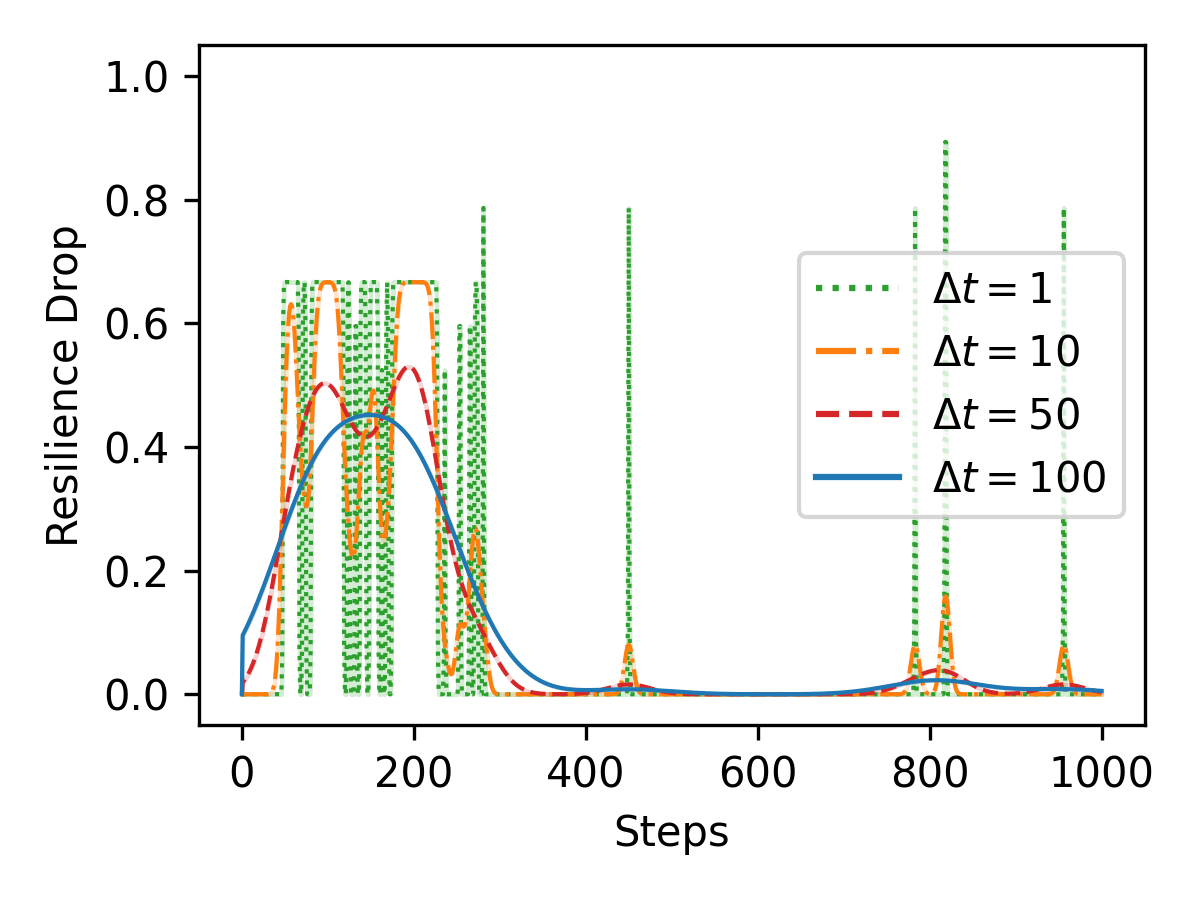}
    \caption{Resilience drop for various application of Gaussian filters with $\sigma = \frac{\Delta t}{2}$. Here $\Delta t$ influences the number of neighboring points considered. Higher $\sigma$ leads to more smoothing, and as such, we can control the trade-off between more detailed information by using smaller  $\Delta t$ and clearer attack shapes using larger $\Delta t$.}
    \label{fig:window_comparison}
\end{figure}

We illustrate the effect of time granularity on the resilience drop metric in Figure~\ref{fig:window_comparison}, using an RL game between a PPO-trained blue agent and a heuristic red agent (details in Section~\ref{sec:rl_defenses}. We use a Gaussian-based rolling mean to smooth the time-series data. This approach applies a one-dimensional Gaussian filter, focusing on data points nearest in time while preserving trends and reducing high-frequency noise (\cref{fig:window_comparison}).
As $\Delta t$ increases, there is a trade-off between information loss and a clearer representation of the attack trend. In this case, the attack ramps up early in the game, and then the blue agent is able to recover and mitigate the adversary's impact. 
When using $\Delta t=1$,  it is challenging to identify patterns within the attack. Through our experiments, we found that $\Delta t = \frac{T}{10}$ allows for swift analysis of an attack pattern, helping to identify general patterns before reducing the window size for attacks that require further inspection.

A clear assessment of cyber resilience needs to support different levels of aggregation across various settings, and facilitate the direct comparison of different defense strategies (properties P1 and P4 from the problem formulation in~\Cref{sec:problem}).
To support these properties, we normalize the resilience drop by dividing it to the maximum possible value drop per interval:
\begin{equation}
\Delta R_{norm} (B,R, \Delta t) = \Delta R (B,R, \Delta t)  / \Delta R_{max}
\end{equation}
The maximum drop per interval $\Delta R_{max}$ occurs when the attacker is successful within every individual step, for a total of $N = \Delta t$ impacts directed at the server with the highest impact cost:
\begin{equation}
%\Delta R_{norm} = \Delta R / \Delta R_{max}\\
\Delta R_{max} = \sum_{i \in \{C,\ A,\ I\}} w_i \Delta t  \max_{j\in Assets} cost(i, j)
\end{equation}
Thus, through normalization, we can aggregate the resilience metric and compare defenses across topologies and attacks to understand the performance of an agent under various conditions (discussed in \Cref{sec:summarization}).

\subsection{Case study: Employee payroll workflow} 
\label{sec:case_study}
We consider an employee payroll workflow, where essential services include submitting work hours, retrieving documents, and various daily operations. At a minimum, three critical servers are present in the network: the authentication server $AS$, a database server $DS$, and a front web server $WS$. Given the three operational goals of confidentiality $C$, availability $A$ and integrity $I$, we obtain the following formula for resilience drop due to potential cyber threats:
\begin{align}
\Delta R (B, R, \Delta t) = \sum_{i \in \{C,\ A,\ I\}} w_i \sum_{j \in \{AS, DS, WS\}} \left(\sum_{t \in \Delta t} N_j(t) \right) \times  cost(i, j)
\end{align}
The resilience decrease unifies the performance drop for the three objectives: $\Delta C$ -- the confidentiality drop due to exfiltration of credentials and user records,  $\Delta A$ -- the availability decrease due to disruption of service for users, and $\Delta I$ -- the integrity drop due to unauthorized website changes or data corruption:
\begin{align}
\Delta R (B, R, \Delta t) = w_C \Delta C (B, R, \Delta t) + w_A \Delta A (B, R, \Delta t) + w_I \Delta I (B, R, \Delta t)
\end{align}
An attacker targeting confidentiality (data theft) impacts the authentication and the database servers, but not the  web server, hence the  confidentiality drop can be defined as:
\begin{align}
\Delta C (B, R, \Delta t) = \sum_{j\in \{AS, DS\}} \left(\sum_{t \in \Delta t} N_j(t) \right) \times cost(C, j)
\end{align}
Similarly, we can define the  availability and integrity drop, as a function of number of impacts and cost on the associated critical services. Through this formulation, we can adapt the objectives to the use cases and tailor the costs to model the scenarios we are interested in. For example, in emergency response systems or critical infrastructure, where uninterrupted service is crucial, the availability objective and associated services will be modeled with weights and impact costs higher than those of other objectives or services.
In corporate settings, sensitive data like proprietary algorithms and private customer information require increased protection to prevent industrial espionage and data theft, and can be modeled by increasing the cost associated with the confidentiality objective.

\subsection{Resilient RL defenses}
\label{sec:rl_defenses}

A resilient defense strategy requires both effective precautions (network hardening) and the ability to recover quickly from an attack. An autonomous agent must be able to absorb the attack quickly and with minimal impact on operations. Thus, a resilient blue agent incorporates the following features:
\begin{enumerate}
\item \emph{Adaptive}: Autonomous agents must be trained against multiple adversarial behaviors and topologies to ensure that they can adapt to diverse settings.
\item \emph{Reactive}: Agents need to react quickly to evidence of compromise in the network, as soon as it is discovered. Specifically, the presence of indicators of compromise (IOCs) in the network, such as malicious files on a host or communication with known malicious IP addresses, should trigger immediate recovery actions. This is achieved by prioritizing recovery actions over other actions when IOCs are present in the agent's observation.
\item \emph{Proactive}: Blue agents should prioritize actions that protect or harden the network (such as setting up decoys) to prepare for incoming attacks. 
\end{enumerate}

Motivated by these principles, we developed several agents to investigate the contribution of each of the above characteristics to the success of the defense strategy.

\begin{itemize}
\item \textbf{PPO} (adaptive): Our blue agent trained with PPO, after hyper-parameter tuning. 
\item \textbf{Blue-R} (adaptive and reactive): A PPO-based agent trained with the same hyperparameters, which also features quick reaction to indicators of compromise in the network. Blue-R uses the Analyse action to determine if a host has been compromised. If so, Blue-R uses action masking to guide the PPO algorithm to choose the next action only from recovery actions such as Restore or Remove, on compromised machines. 
\item \textbf{Blue-RD} (adaptive, reactive and proactive): An enhancement of Blue-R, which also uses Deploy Decoy action to harden hosts in the network proactively. This action strategically places fake services to lure attackers away from real operational services. The blue agent detects the activity of the attacker when the red agent interacts with the decoy service. If there are no indicators of compromise on the network, the blue agent prioritizes setting up fake services with the goal of having at least one decoy active on each host. Blue-RD also keeps track of all services to avoid any attempts to set up decoys on ports that are used by normal services.
\end{itemize}

\section{Experimental Results}
\label{sec:single_attack}

We ran the experiments on an extension of the CybORG Cage 2 framework, which allows the generation of random network topologies consisting of 3 or 4 subnets, with 2 to 5 hosts per subnet, with a similar network setup as described in Figure~\ref{fig:network}. One of the subnets includes three critical servers that support system operations: an authentication, a database and a front web server.
CybORG comes with a series of red and blue heuristic agents that can be used in testing. In our experiments, we used the CybORG's strongest heuristic randomized red agent, B-line, which scans the hosts from its list of known hosts at random, looking for vulnerabilities, and attempts to reach the critical services following the shortest path. We also used two heuristic blue agents (Monitor and Restore) as baselines to compare with the RL blue agents that we developed and trained.

\subsection{Comparing defense strategies}

\begin{figure}
\centering
\begin{subfigure}{.4\textwidth}
  \centering
  \includegraphics[width=1\linewidth]{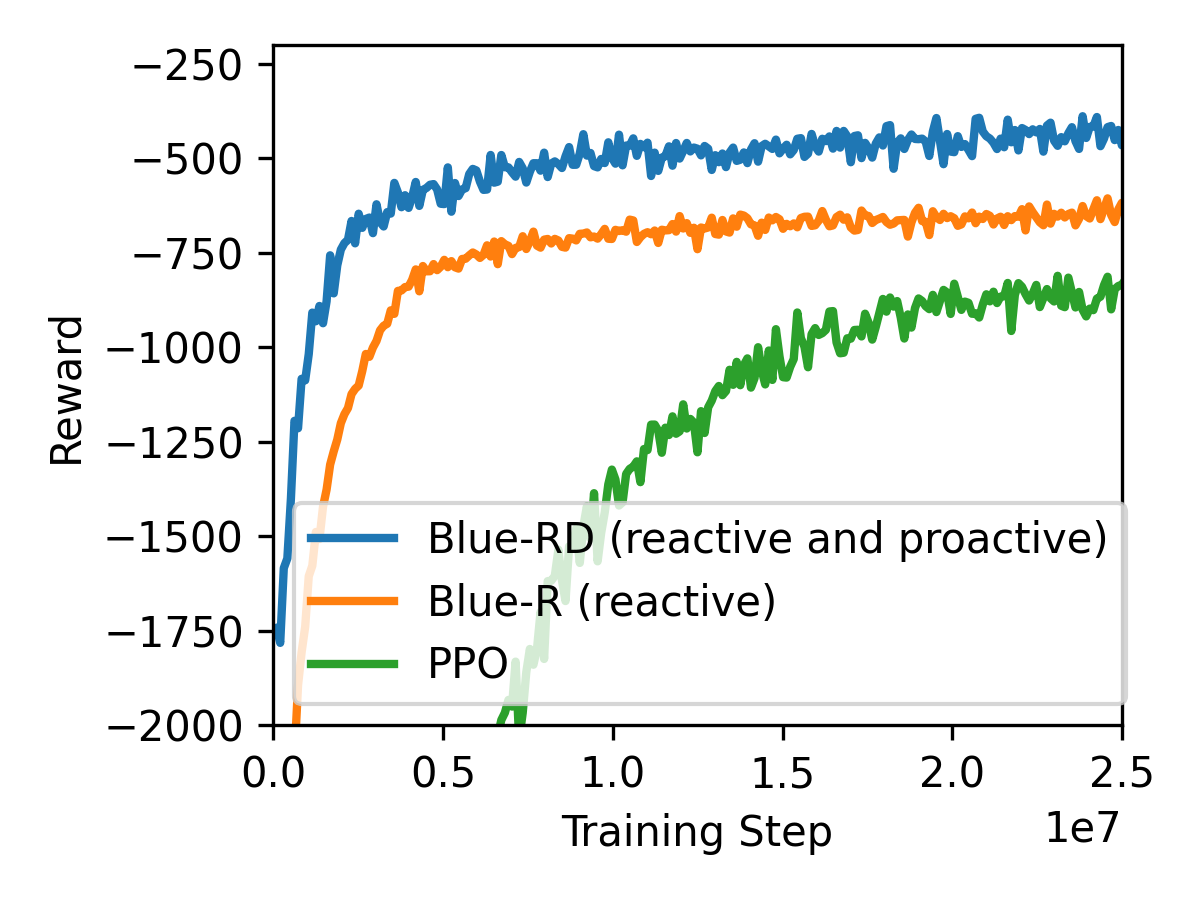}
  \caption{Training (Reward)}
  \label{fig:sub1}
\end{subfigure}%
\begin{subfigure}{.4\textwidth}
  \centering
  \includegraphics[width=1\linewidth]{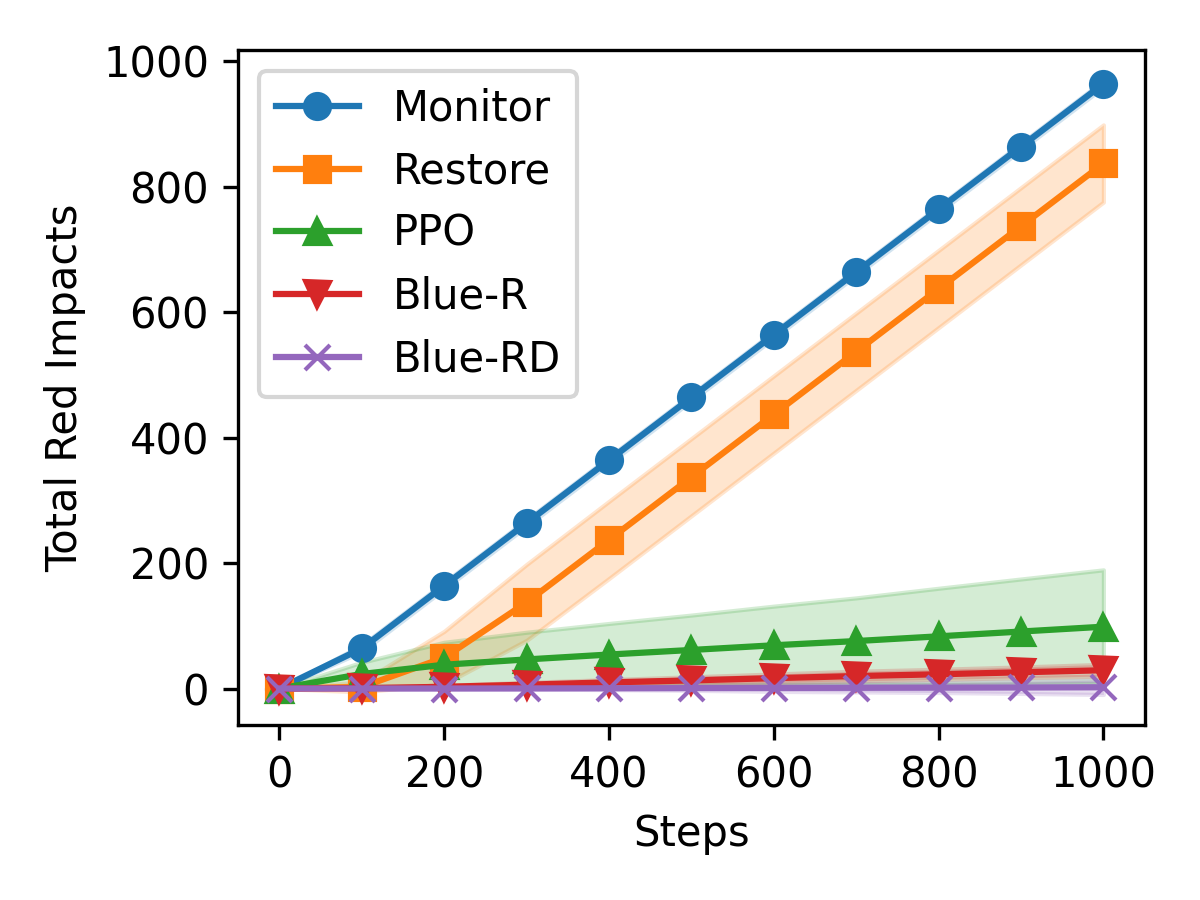}
  \caption{Evaluation (Total Impacts)}
  \label{fig:sub2}
\end{subfigure}
\caption{Comparing blue agents. (a) Reward during training for the three PPO-based agents. (b) Total number of successful adversarial impacts during evaluation, using trained RL models, and the two heuristic baselines (averaged over 100 episodes).}
\label{fig:compare_defenses}
\end{figure}

We compare the RL agents presented in \Cref{sec:rl_defenses} with the CybORG heuristic agents in Figure~\ref{fig:compare_defenses}. Figure~\ref{fig:sub1} shows the reward during training for the three trained defenses, PPO, Blue-R, and Blue-RD. The blue agent is penalized (by -0.1) when the red agent is able to gain root access to a hosts, when the red agent is successful in calling the Impact action on a critical server (by -10), and when a machine is restored (by -1). The maximum possible reward for the blue agent is zero. Both Blue-R and Blue-RD converge faster than the basic PPO strategy, because they use action space masking to guide the defense in choosing from a tailored subset of actions based on the presence of indicators of compromise.  
Figure~\ref{fig:sub2} presents how successful the attacker is against each defender by counting the number of Impacts during evaluation and averaging it over 100 episodes. The three trained defenses perform significantly better than the baseline rule-based blue agents provided in CybORG (Monitor and Restore). The Blue-RD defense is able to mitigate the attack the most, as it employs both proactive security measures and a fast response to compromise. At the end of the game, the network faced 2, 29, 99, 837, and 964 adversarial impacts under the Blue-RD, Blue-R, PPO, Restore, and Monitor defenses, respectively (100-episode averages).

\subsection{Resilience of the system under attack}
\begin{figure}[t!]
    \centering
    \includegraphics[width=1\linewidth]{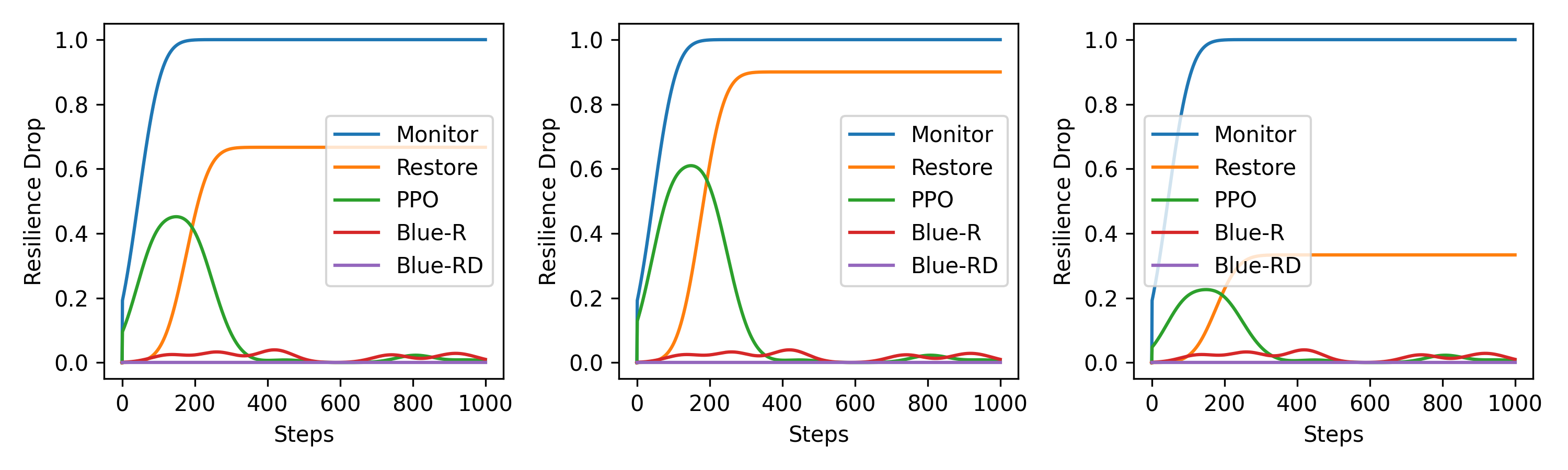}
    \caption{Resilience drop for each blue agent during the same attack. (Left) goals and assets are equally important (Weights-1, Costs-1). (Center) availability of resources is prioritized (Weights-2, Costs-1). (Right) authentication server is ranked the most critical resource (Weights-1, Costs-2). }
    \label{fig:single}
\end{figure}
In Figure~\ref{fig:compare_defenses}, we have compared the defense strategies based on episodic return and the number of adversarial impacts on critical servers. However, these cumulative metrics do not capture how the resilience of the system has evolved during the attack. 
In this section, we are specifically investigating the evolution of resilience in the context of \emph{a single attack}, while in Section~\ref{sec:summarization} we will discuss how to understand and evaluate the resilience of a system over multiple attacks and network topologies. We selected one of the 100 episodes averaged in Figure~\ref{fig:sub2} to evaluate the resilience metric during the course of an attack. 
A similar analysis can be applied to any other episode. 
 
Weights are used to balance the operational goals of confidentiality, availability, and integrity, while costs guide the criticality of each asset (authentication, database, and front web server) per operational goal. 
In this analysis, we use the sets of weights and costs described next.

\begin{itemize}
\item \textbf{Weights-1:} Equal importance for the three operational goals: $w_C = w_A = w_I = 1/3$.
\item \textbf{Weights-2:} Higher importance for availability: $w_A = 0.8$; $w_C = w_I = 0.1$.
\end{itemize}
\begin{itemize}
\item \textbf{Costs-1:} Same cost for all assets relevant to each goal. 
\item \textbf{Costs-2:} Different costs per asset. The authentication server is considered the most critical and an impact cost is assigned that is $2\times$ higher than the other servers. 
\end{itemize}

We explore the following research questions in Figure~\ref{fig:single}: 

\paragraph{\textbf{(Q1) How do different defenses compare in terms of resilience?}} 
The ability to compare defenses in a repeatable and verifiable way (Property P4 from \Cref{sec:problem}) is a necessary feature for a resilience metric, as it informs what security measures must be implemented in the network.  
The graph in Figure~\ref{fig:single}-Left quantifies the drop in resilience when all operational goals and relevant assets are equally important (Weights-1, Costs-1). Although the ranking of defenses is the same as before (Figure~\ref{fig:sub2}), thus reinforcing the findings that both proactive and reactive security measures are needed to maintain operational workflows, the plot provides new insights about the evolution of the system under attack. 

Property P2 of the metric, specifically the ability to capture time-dependent evolution patterns, is crucial in understanding when the attack starts, how soon it is contained, and whether the system is able to absorb the attack, bounce back, or simply collapse and never recover. 
The Monitor blue agent, which only collects alerts, but does not actively defend the network, incurs the largest drop in resilience; once the attacker has reached an essential service (i.e., the authentication server) it will keep using the Impact action to collect rewards. The Restore blue agent cleans out some of the hosts that present suspicious processes, but, eventually, the red agent successfully exploits and impacts one of the critical services, and the Restore agent is not able to regain functionality of that service.
The PPO blue agent experiences a strong attack within the first part of the game but is able to learn to recover and prevent future attacks from escalating. The Blue-R successfully mitigates the attack with timely recovery actions that prevent it from spreading, while the Blue-RD strategy performs the best, fully maintaining the system operations by proactively hardening hosts.

\paragraph{\textbf{(Q2) How does the resilience drop depend on operational goals?}}
Emergency response systems or critical infrastructure are just some of the cases where the need for uninterrupted service outweighs other concerns. In Figure~\ref{fig:single}-Center, we study a situation where the availability of resources is more important than other operational goals (Weights-2, Costs-1). 
During the attack studied here, the PPO and Restore defenses fail to protect essential services (the front web server and the database, respectively), which leads to a decrease in resilience. 

Note that although the attack is the same as in Figure~\ref{fig:single}-Left (same number of adversarial impacts) prioritizing availability makes the same disruptions carry more weight. 
In the case of the PPO defense, for example, the largest service interruption in the network occurs on the front web server (126 total impacts). For e-commerce platforms, which rely heavily on online operations, the availability of the front web server is crucial to prevent revenue loss. Hence, adjusting the weights on operational goals accordingly helps defenders correctly assess the scale of the problem and employ the most effective defenses. In this case, securing the front web server should be the highest priority for defenders. 

\paragraph{\textbf{(Q3) How does the resilience drop depend on the importance of different services?}} Observing which machines are driving down the resilience of the system can inform specific security measures to prevent or mitigate attacks. However, such measures often require an investment in redundant equipment, security software or human labor and can increase the financial costs of maintaining operations.
To limit these costs, it is necessary to understand the degree of impact that various network components have on the resilience of the system.

In Figure~\ref{fig:single}-Right, we present a situation where the authentication server is the most critical resource in the network (Weights-1, Costs-2). This is the case in various domains like healthcare, banking, finance, where authentication is essentially the first line of defense against unauthorized access to sensitive data and systems. As a central network resource for security management, an authentication server must run special and usually expensive software~\citep{SHINDER2008121}. 

During the attack investigated here, PPO and Restore have a difficult time protecting the front web server and the database, but they are both effective at securing the authentication server. Therefore, the decrease in resilience for PPO and Restore in Figure~\ref{fig:single}-Right, where the authentication server is crucial, is smaller than in Figure~\ref{fig:single}-Left, where all services were ranked equally important. Whether this level of resilience is sufficient depends on the real-world application. 
Nevertheless, using a metric that can prioritize objectives and services according to operational goals is essential to help balance the cost effectiveness of security solutions with long-term resilience goals.

\section{Quantifying System Resilience over Multiple Attacks and Topologies} \label{sec:summarization}

In \Cref{sec:single_attack}, we evaluated the resilience of a blue agent for a fixed red agent (attacker) and network topology. Here, we first discuss several methods to summarize information over multiple runs of a game (\Cref{sec:info_summary}). Then we apply these methods to evaluate the resilience of the system across multiple attacks on the same topology (\Cref{sec:multiple_attacks_single_topology}) and on various topologies (\Cref{sec:multiple_attacks_multiple_topologies}). Note that the aggregation property of the resilience metric (Property P1 from Section \ref{sec:problem}) is essential to understand the resilience of the system in diverse settings of interest. 

\subsection{Summarizing Information From Multiple Games} 
\label{sec:info_summary}

Let $S$ be the total number of game steps and $N$ the number of games. Each game essentially includes a different randomized attack, controlled by varying the random seed of the environment. 
Each game can be represented as a vector of resilience drops $\mathbf{r_{\Delta{t}}} = [r_{1_{\Delta{t}}}, \cdots, \cdots r_{T_{\Delta{t}}}] \in \mathbb{R}^{T}$ where $T = \lfloor \frac{S}{\Delta{t}}\rfloor$; $r_{k_{\Delta{t}}}$ represents the value of $\Delta{R}$ during the $k$-th time interval $\Delta t$.
From $\mathbf{r_{i_{\Delta{t}}}}$ of all attack vectors $i\in \{1, \cdots, N\}$ we can build the resilience drop matrix $\mathbf{R}^{N \times T}$.

\paragraph{\textbf{Fine-Grained View: Individual Attacks.}}
We can inspect the resilience of a blue agent for individual attacks; however, while each attack provides us with a fine-grained view of the resilience, it does not facilitate a comparison of defenses over multiple settings. 

\paragraph{\textbf{Coarse View: Averages and Standard Deviations.}}
We use $\mathbf{R}$ to calculate the mean resilience drops as $\mathbf{\mu} = \frac{1}{N}\mathbf{1}^{T}\mathbf{R} \in \mathbb{R}^{1 \times T}$, from which we obtain the centered matrix of resilience drop $\tilde{\mathbf{R}} = \mathbf{R} - \mathbf{1\mu} \in \mathbb{R}^{N \times T}$, where $\mathbf{1} = [1, \cdots, 1]^T \in \mathbb{R}^{N \times 1}$. We then obtain the standard deviation of the resilience drop as $\sigma = \sqrt{\frac{1}{N}\mathbf{1}^T(\tilde{\mathbf{R}} \odot \tilde{\mathbf{R}})} \in \mathbb{R}^{1 \times T}$ where $\odot$ is the \citet{Hadamard}.
We use $\mu$ and $\sigma$ to obtain an overview of the resilience of an agent.
Although this approach gives us a summary of the resilience of an agent, it provides a coarse view and might miss variations of resilience patterns. 

\paragraph{\textbf{Balanced View: Clustering Attacks.}} 
Instead of taking the average over all attacks, we can first apply a clustering algorithm to group attacks that share patterns.
We compute the pairwise distance matrix $\mathbf{D}$  such that $D_{i,j} = d(\mathbf{r_{i}}, \mathbf{r_{j}})$ using the Euclidean distance $d(\mathbf{r_i}, \mathbf{r_j}) = ||\mathbf{r_i} - \mathbf{r_j}||_2$.
We perform an agglomerative clustering using Ward's method \citep{wardHierarchicalGroupingOptimize1963} based on the distance matrix, with a fixed number of clusters ($K=3$). 
These clusters partition $\mathbf{R}$ into subsets of rows, which we interpret as $K$ resilience drop matrices $\{\mathbf{R_k}\}_{k \in {1, \cdots,K}}$ where $\mathbf{R_k}\in \mathbb{R}^{N_{k} \times T}$ represents the resilience drop matrix associated with the $k$-th cluster and $N_k$ is the number of attacks within cluster $k$. We then compute their respective means $\mu_{k}$ and standard deviations $\sigma_k$.

\subsection{Resilience for multiple attacks on the same topology} 
\label{sec:multiple_attacks_single_topology}
\begin{figure}[t!]
    \centering
    \includegraphics[width=1.0\linewidth]{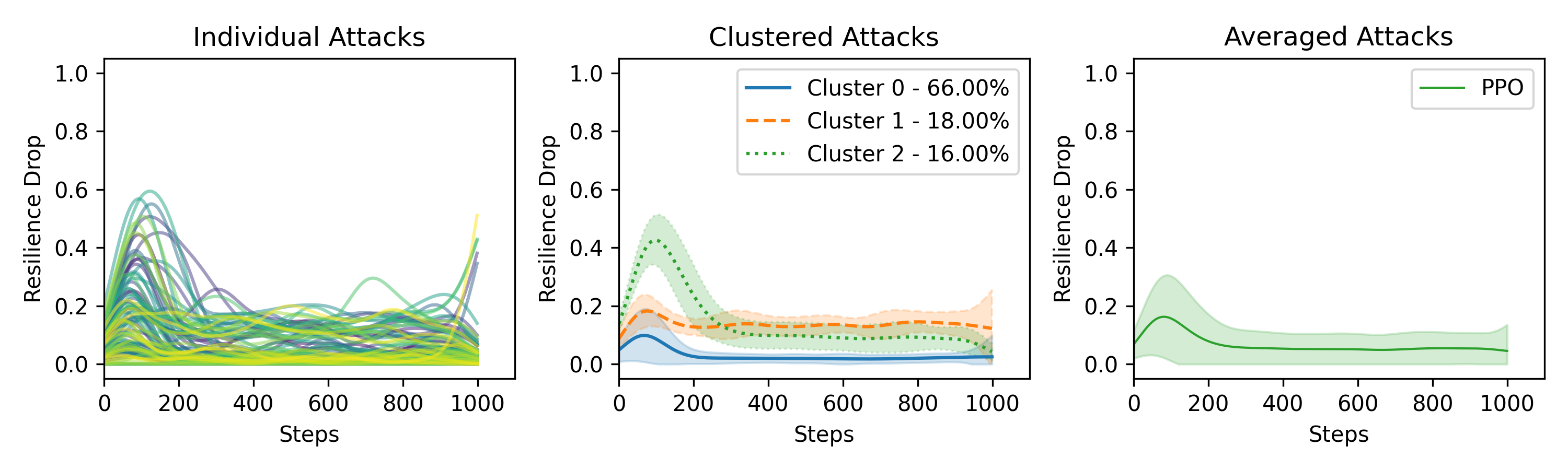}
    \caption{Resilience drop for PPO blue agent over multiple attacks on the same topology. (Left) Each attack is shown individually. (Center) Attacks are Clustered into $K=3$ clusters. (Right) Attacks are summarized using the average resilience decrease and the standard deviation.}
    \label{fig:mast_views}
\end{figure}
We present several experiments using the PPO agent, with the goal of characterizing the global resilience of the system when faced with multiple attack patterns on the same network topology. 
We ran 100 different games on a fixed topology, each game having a different random seed and thus, representing a different attack. We used a time interval $\Delta t$ of 100 to visualize the resilience drop across time. 
\cref{fig:mast_views}-Left shows the decrease in resilience during individual runs. 
\cref{fig:mast_views}-Right presents a coarse-grained average resilience drop over all 100 attacks. Note that the mean resilience follows the shape of single-attack resilience curves. Simply by inspecting the mean resilience curve, one can tell that the adversary is usually successful in ramping up an attack against the PPO blue agent, but, eventually, the blue agent recovers and is able to mitigate the attack. 
\cref{fig:mast_views}-Center partitions attacks into three clusters based on the resilience patterns over the course of the games. The blue agent is able to recover partially after 300 steps and mitigate the attack, but for 16\% of the attacks, resilience decreases twice as much as the overall average (note the 0.4 peak of Cluster 2 compared to 0.2 in the rightmost graph). Security operators can use this information as feedback to investigate attack patterns from Cluster 2 and incorporate additional defenses to better handle and adapt to this group of attacks.

\subsection{Resilience for multiple attacks on various topologies} \label{sec:multiple_attacks_multiple_topologies}

We consider a set $S$ of topologies, where $|S|$ is the number of topologies in the set.
Let $\mathbf{R}^{(s)} \in \mathbb{R}^{N \times T}$ be the resilience drop matrix of a blue agent over $N$ attacks on a given topology $s$. For the set of topologies, we have an associated set of resilience drop matrices $\{\mathbf{R}^{(s)}\}_{s \in S}$ of size $|S|$.
We build $\mathbf{R}^{total}$ as the concatenation of $\mathbf{R}^{(s)}$, specifically $\mathbf{R}^{total} = [\cdots, \mathbf{R}^{(s)}, \cdots]^T \in {\mathbb{R}^{(N \times |S|) \times T}}$.
We can then apply the summarization methods discussed before to $\mathbf{R}^{total}$ and evaluate the resilience of the blue agents over multiple topologies and attacks.

\begin{figure}[t!]
    \centering
    \includegraphics[width=1.0\textwidth]{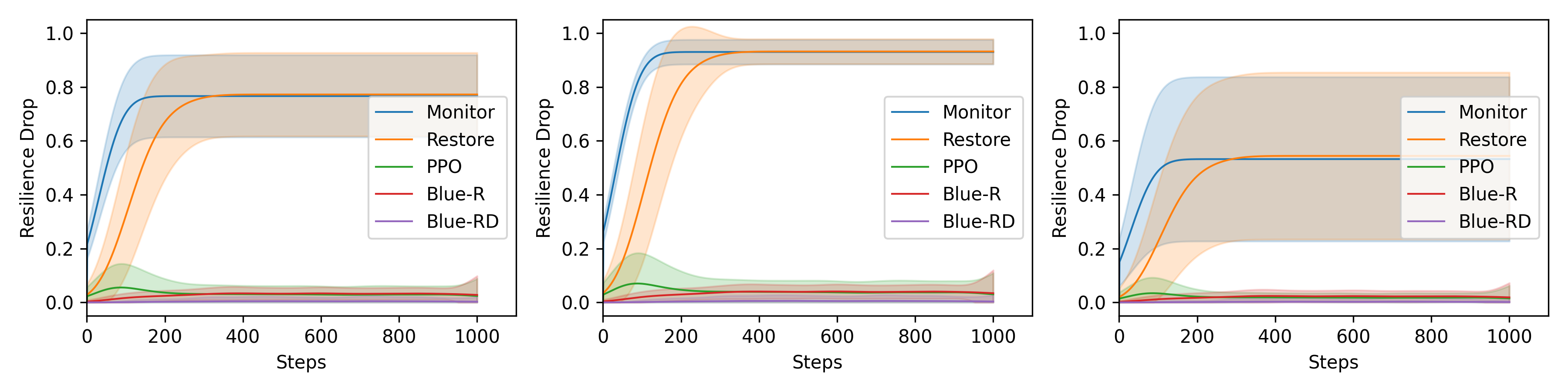}
    \caption{Resilience drop averaged over 5 different network topologies and $100$ attacks per topology. 
    (Left) goals and assets are equally important (Weights-1, Costs-1). (Center) availability of resources is prioritized (Weights-2, Costs-1). (Right) authentication server is ranked the most critical resource (Weights-1, Costs-2). 
    }
    \label{fig:mamt_random_selected}
\end{figure}

Using this approach, we compare the five blue agents Monitor, Restore, PPO, Blue-R and Blue-RD over five different topologies $|S|=5$, and $N=100$ attacks on each topology. From \cref{fig:mamt_random_selected}, we see that the ranking of the defenses is consistent with what was observed in ~\Cref{fig:compare_defenses} for a single attack. Blue-RD, the agent that incorporates both proactive and reactive measures, outperforms the other strategies and is able to keep the system resilient across all the settings studied here. Resilience averaging smooths out more extreme variations in individual runs, offering a more conclusive comparison. Counterintuitively, Monitor, the agent that simply observes the network, without taking any action to defend it, does not reach a resilience drop of 1 in all cases.  
The reason is that once the red agent is able to compromise a service, it will choose to impact the same service to collect rewards, if no defense deters it. However, the affected service may be less critical for the operational goals; since the maximum possible resilience drop is not reached, the normalized value will be less than 1.  

\section{Conclusion}
This work introduced a quantitative resilience metric to evaluate autonomous agents for cyber defense.
This metric allows security operators to assess and compare defensive strategies across various attack patterns and network topologies. It can also be adapted to align with specific operational goals and asset criticality.
Using this metric, we demonstrate the value of integrating proactive and reactive defensive measures.
In particular, reinforcement learning-based agents that incorporate network hardening techniques and rapid response mechanisms significantly enhance resilience.
Our framework prioritizes key security objectives—confidentiality, integrity, and availability—and provides actionable insights for optimizing cyber defenses in dynamic threat environments. The resilience metric can also be used to guide the training of defense strategies. However, if the operational goals change, the model needs to be re-trained or fine-tuned. We leave as future work the study of using this metric to learn defense strategies that can adapt with minimal overhead as workflow priorities change.

\subsubsection*{Acknowledgments}
\label{sec:ack}
This research was funded by the Defense Advanced Research Projects Agency (DARPA), under contract W912CG23C0031.

\bibliographystyle{unsrtnat}
\bibliography{main}

\end{document}